\documentclass[twocolumn,prb,aps,amsfonts,amssymb,amsmath,showpacs,floatfix]{revtex4-1}
\usepackage{graphicx}% Include figure files
\usepackage{dcolumn}% Align table columns on decimal point
\usepackage{mathtools}% include equation features
\usepackage{color}
\usepackage{braket}
\usepackage{hyperref}
\graphicspath{{figures/}}
\usepackage{soul}

\begin{document}
\title{Suppressing quasiparticle poisoning with a voltage-controlled filter}

\author{Gerbold~C.~M\'enard$^{1,*}$, Filip~K.~Malinowski$^{1,*}$, Denise Puglia$^{1}$, Dmitry~I.~Pikulin$^{2}$, Torsten~Karzig$^{2}$, Bela~Bauer$^{2}$, Peter~Krogstrup$^{3}$, Charles~M.~Marcus$^{1}$.}

\affiliation{ $^{1}$ Center for Quantum Devices, Niels Bohr Institute, University of Copenhagen and Microsoft Quantum Lab, Universitetsparken 5, 2100 Copenhagen, Denmark\\
$^{2}$ Microsoft Station Q, University of California, Santa Barbara, California 93106-6105, USA\\
$^{3}$ Microsoft Quantum Materials Lab, Kanalvej 7, 2800 Lyngby, Denmark \\
$^{*}$ These authors contributed equally to this work}

\date{\today}

\begin{abstract}
We study single-electron charging events in an Al/InAs nanowire hybrid system with deliberately introduced gapless regions. The occupancy of a Coulomb island is detected using a nearby radio-frequency quantum dot as a charge sensor. We demonstrate that a 1~$\mu$m long gapped segment of the wire can be used to efficiently suppress single electron poisoning of the gapless region and therefore protect the parity of the island while maintaining good electrical contact with a normal lead. In the absence of protection by charging energy, the 1$e$ switching rate can be reduced below $\Gamma = 200$~s$^{-1}$. In the same configuration, we observe strong quantum charge fluctuations due to exchange of electron pairs between the island and the lead. The magnetic field dependence of the poisoning rate yields a zero-field superconducting coherence length of $\xi=90 \pm 10$~nm.
\end{abstract}

\maketitle

\section{Introduction}
Semiconductor-superconductor hybrids combine coherence effects at the macroscopic scale (superconductors) with the ease of tuning by means of electric and magnetic fields (semiconductors). An ever-growing class of phenomena enabled by these hybrids include quantum phase transitions\cite{sachdev2011} such as superconductor-insulator transition\cite{han2014,mooij2015,bottcher2018} or topological superconducting transitions\cite{albrecht2016,vaitiekenas2018_2,mourik2012,das2012,deng2016,nichele2017,menard2017,zhang2018,chen2017}, gate-tunable transmon \cite{larsen2015,delange2015,casparis2017}, and proposed topological Majorana qubits \cite{aasen2016,karzig2017}. Quasiparticle poisoning\cite{aumentado2004,vanwoerkom2015,vanveen2018,zgirski2011,albrecht2017,rainis2012} presents a significant source of decoherence for hybrid and superconducting qubits\cite{martinis2009,catelani2011,sun2012,riste2013,gustavsson2016,mannila2018}. This is especially true due to lower induced gap in the semiconducting part of the heterostructure.

This article describes a method that can significantly reduce the detrimental effects of quasiparticle poisoning. Specifically, we introduce a tunable quasiparticle filter made from an InAs nanowire with an epitaxial aluminum shell on two facets\cite{krogstrup2015,chang2015,gazibegovic2017}. Such a component can provide electrical connection, with Cooper pairs as charge carriers, while keeping quasiparticle transport to a minimum between two segments of the device: one that is poisoned and another that needs to remain quasiparticle-free.

An example of a system that is in need of a quasiparticle filter is a Majorana-based topological qubit\cite{aasen2016,karzig2017}. Proposed designs of such a qubit commonly require large-scale (potentially strongly poisoned) superconducting networks connected to non-superconducting leads and smaller regions of topological superconductor hosting Majorana zero modes, which decohere as a result of the poisoning\cite{goldstein2011,rainis2012,budih2012}.

The devices under study consist of two gapless regions (metallic lead and soft-gapped proximitized nanowire) separated by a clean InAs nanowire segment with an epitaxial aluminum shell~\cite{krogstrup2015,chang2015,gazibegovic2017} with a tunable gap~\cite{mikkelsen2018,antipov2018}. To enable charge detection, the clean InAs/Al and soft-gapped regions are configured as a single island (i.e., quantum dot). By means of radio-frequency charge sensing with microsecond temporal resolution\cite{jung2012,petersson2010,barthel2010,DR2018,HQN2018}, we observe single-electron tunneling events between two zero-gapped regions while tuning the superconducting gap of the filter by electrostatic gating and applied magnetic field. Increasing the coupling of the semiconductor part of the clean nanowire segment to the superconductor, resulting in the hard gap, suppresses the single-electron tunneling events between the zero-gapped regions. At the same time, the island remains strongly coupled to the lead as revealed by observed quantum charge fluctuations.

This article is organized as follows. Section~\ref{device_and_methods} presents the description of the devices and methods. Section~\ref{DC_characterization} is dedicated to the characterization of the device using conventional lock-in techniques and radio-frequency measurements. Section~\ref{RF_charge_diagram} introduces charge stability diagrams of the studied island. Section~\ref{method} describes specific measurement and analysis protocol used to quantify the quasiparticle poisoning rate. In Section~\ref{efficiency} we use these protocols to perform analysis of the quasiparticle poisoning rate as a function of gate voltages and external magnetic field. Section~\ref{fluctuations} demonstrates evidence of quantum charge fluctuations in the device configuration characterized by low poisoning rate. Finally, summary of findings and potential applications of the quasiparticle nanowire filter are presented in Section~\ref{conclusions}.

\section{Device and experimental methods}
\label{device_and_methods}

\begin{figure}[bt]
	\includegraphics[width=0.48\textwidth]{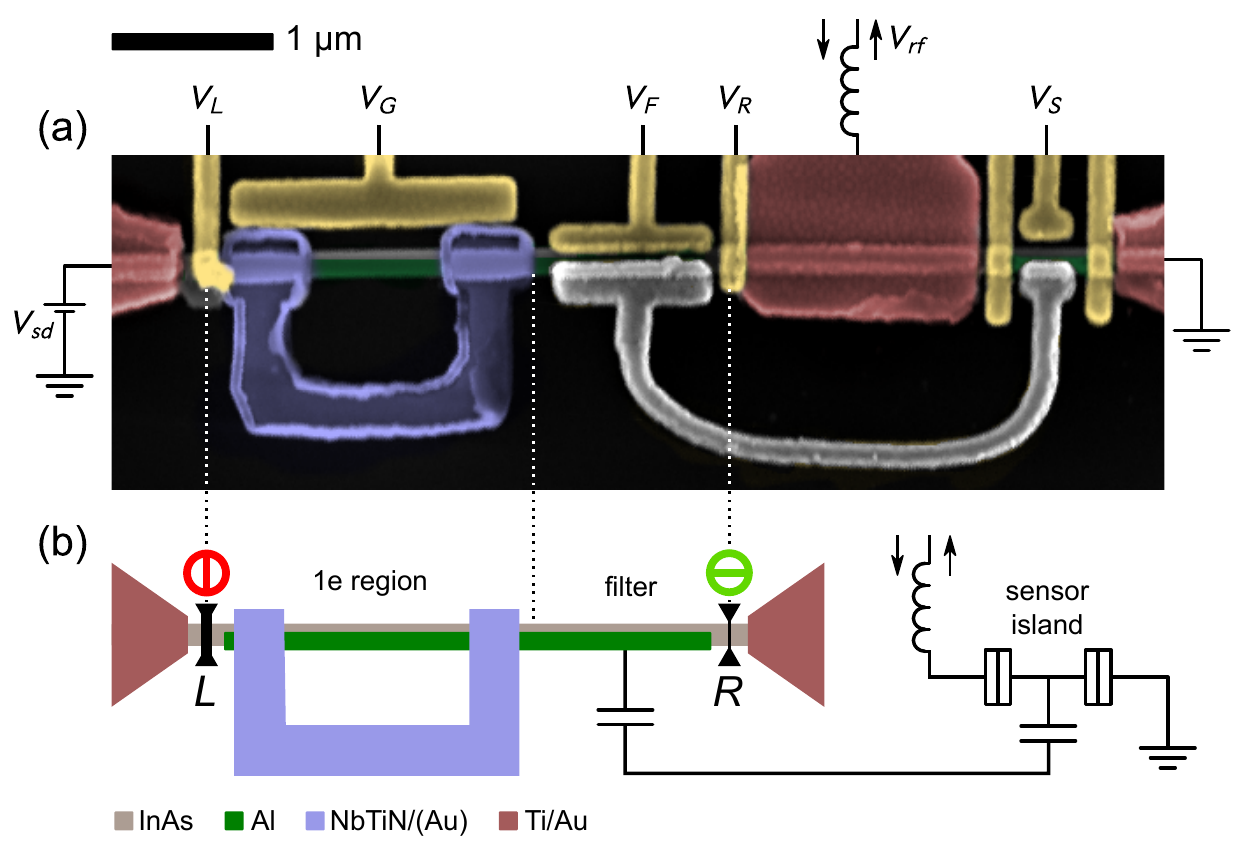}
    \caption{(a) Colorized scanning electron micrograph of one of the devices under investigation. The devices are based on an InAs nanowire (dark-grey) with epitaxial aluminum shell on two of the six facets (green). The aluminum shell is nominally 5~nm thick, but due to oxidation in contact with air we expect it to be effectively 2-3~nm thick. Red-colored regions are Ti/Au ohmic contacts. Gates are used to define and control occupancy of the two islands (yellow). NbTiN/(Au) structure (blue) emulates complex high-field--compatible superconducting network. Capacitive coupling between the islands is enhanced by a floating metallic bridge (white-colored). (b) Simplified schematic of the device. During the key measurements the barrier connecting the large island to the lead on the left side \emph{L} is fully pinched-off preventing any tunneling events on a minute timescale, meanwhile the right barrier \emph{R} is operated in the tunneling regime. The smaller island is operated as a rf charge sensor. The 40~MHz LC resonant circuit is indicated by a coil symbol.}
    \label{device}
\end{figure}

The study was performed on two lithographically similar devices, illustrated in Fig.~\ref{device}. In the following, we specify in the caption of each figure for which of the two devices (A or B) the corresponding data set was obtained. We first describe their structure from the fabrication perspective and later discuss the purpose that each element serves in our experiment.

The devices are based on an InAs nanowire with a 2~nm thick MBE-grown Al shell on two facets\cite{krogstrup2015}. The nanowire is placed on a SiO$_2$ insulating substrate, and the metallic contacts and gates are deposited using e-beam lithography. The nanowire is contacted by three Ti/Au leads (colored in red in Fig.~\ref{device}(a)), deposited after local wet etching of the Al shell using Transcene D and Ar milling to obtain ohmic contact with the semiconductor. Blue-colored 150~nm NbTiN is in electrical contact with the wire. In case of device A, NbTiN is covered by 5~nm of gold, intended as a quasiparticle trap. The nanowire, ohmic contacts, and NbTiN/(Au) structure are covered by 7~nm of ALD-grown HfO$_2$. We deposit additional Ti/Au gates on top of the oxide layer. The voltage on yellow-colored gates is controlled, while the gray-colored gate is electrically floating.

The device consists of two islands defined in the Al-covered nanowire. In the experiment, we study 1$e$ charging rates of the larger, left island, detected using the small, right island as a charge sensor\cite{jung2012,barthel2010,DR2018,HQN2018}. Charge sensing is enhanced by means of a floating gate which increases the capacitive coupling between the two islands.

The main (left) island consists of two different parts. The first one, labeled ``1$e$ region'' in Fig.~\ref{device}(b), is connected to the blue-colored NbTiN structure. This segment is gapless due to at least one of these effects: the choice of the gate voltage $V_G$ for which weakly-proximitized states in the semiconductor are highly populated\cite{antipov2018}, the intrinsic softness of the superconducting gap observed in hybrid InAs/NbTiN and Al/NbTiN structures\cite{vanwoerkom2015,gul2018} and, in case of device A, gold quasiparticle traps evaporated on top of the NbTiN structure.

The second part of the quantum dot, labeled ``filter'' in Fig. \ref{device}(b), is a nanowire covered by epitaxial Al. By applying a voltage $V_F$ to the neighboring gate, the coupling between the semiconducting wire and the superconducting shell in the segment can be tuned, effectively tuning a subgap density of states\cite{vaitiekenas2018,mikkelsen2018,antipov2018}. The Al shell of the nanowire is continuous throughout the island, which prevents the creation of an unintentional barrier dividing a single island in two.

The main island (1$e$ region and filter) is connected via gateable barriers to two normal-metal leads, one neighboring the 1$e$ region, the other neighboring the filter. Adjusting barrier gate voltages $V_L$ and $V_R$ allows us to tune the barriers to the two leads from open regime (conductance through the barrier $>$~2~$e^2/h$), to tunneling regime (conductance $\simeq$~0.5~$e^2/h$), to fully closed (tunneling times of minutes or longer).

Measurements are performed using several techniques. The differential conductance through the main island can be measured using standard lock-in techniques. Alternatively, a radio-frequency resonant circuit connected to the central lead (indicated by a coil symbol in Fig.~\ref{device}) is also used in order to perform an effective differential conductance measurement through the main or sensor island, by using rf reflectometry combined with analog homodyne demodulation. When conductance through the sensor island is suppressed, reflectometry can be used as a substitute for lock-in measurements\cite{DR2018}. Finally, when conductance through the main island is suppressed, and the sensor island is tuned to the slope of the Coulomb peak, the charge on the main island can be detected with microsecond temporal resolution\cite{DR2018}.

The experiment was performed in a dilution refrigerator equipped with a vector superconducting magnet. The mixing chamber plate of the refrigerator was at base temperature of 20~mK and, unless mentioned otherwise, no magnetic field was applied.

\section{Low frequency characterization}
\label{DC_characterization}

\begin{figure}[tb]
    \includegraphics[width=0.48\textwidth]{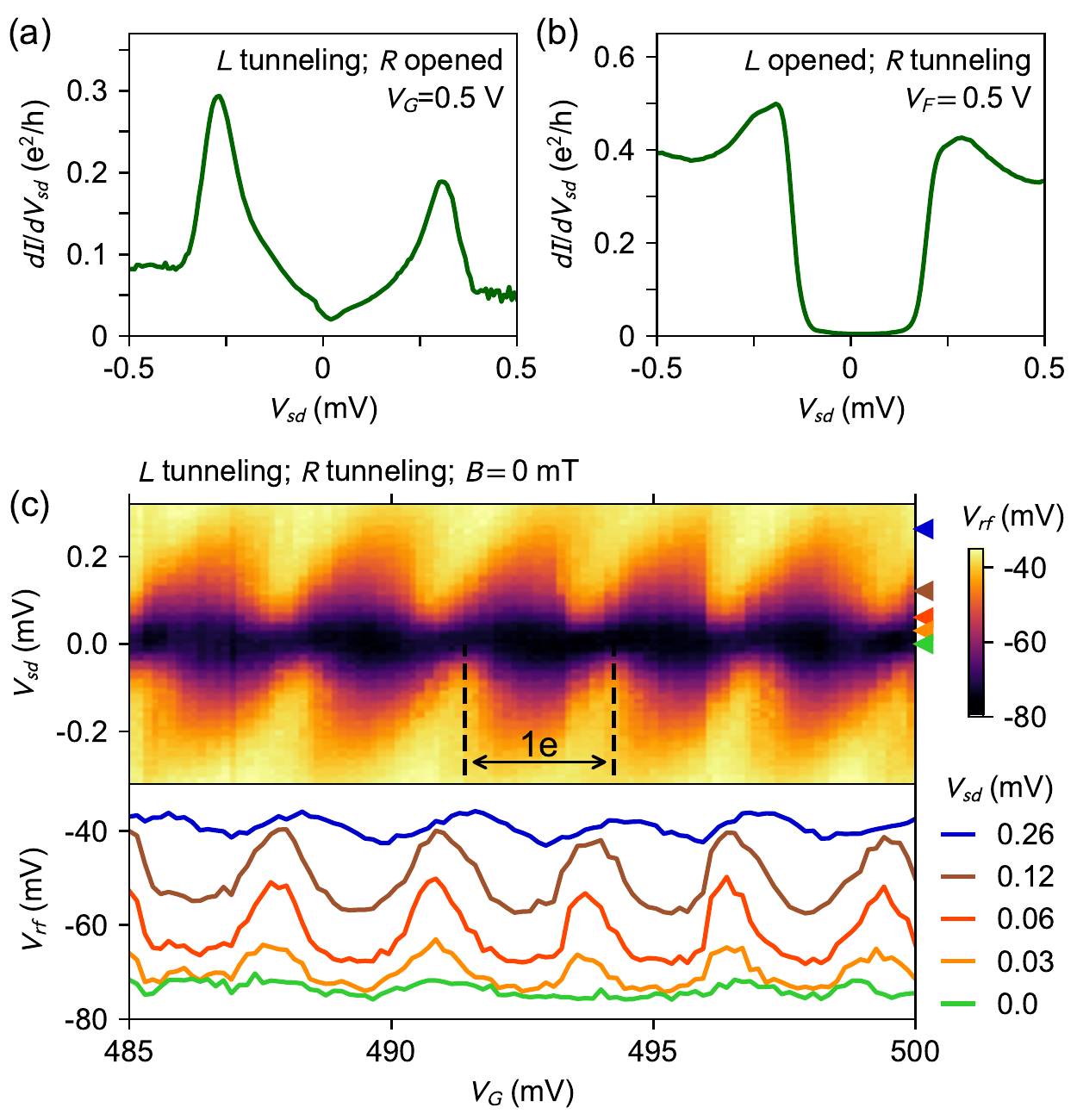}
    \caption{Tunneling spectroscopy of the superconducting gap at the left (a) and right (b) end of the main island obtained using lockin techniques. The inset text indicate the status of the left (L) and right (R) barriers. (c) Coulomb diamonds with very strongly suppressed zero bias conductance obtained by reflectometry measurements. Bottom panel shows cuts through the data at various $V_{sd}$. Data for device A.}
    \label{DC_superconducting}
\end{figure}

We first characterize the main island in our device using standard lock-in techniques or equivalent reflectometry measurements\cite{DR2018}. Figure~\ref{DC_superconducting} summarizes our findings for zero external magnetic field.

\begin{figure}[tb]
    \includegraphics[width=0.48\textwidth]{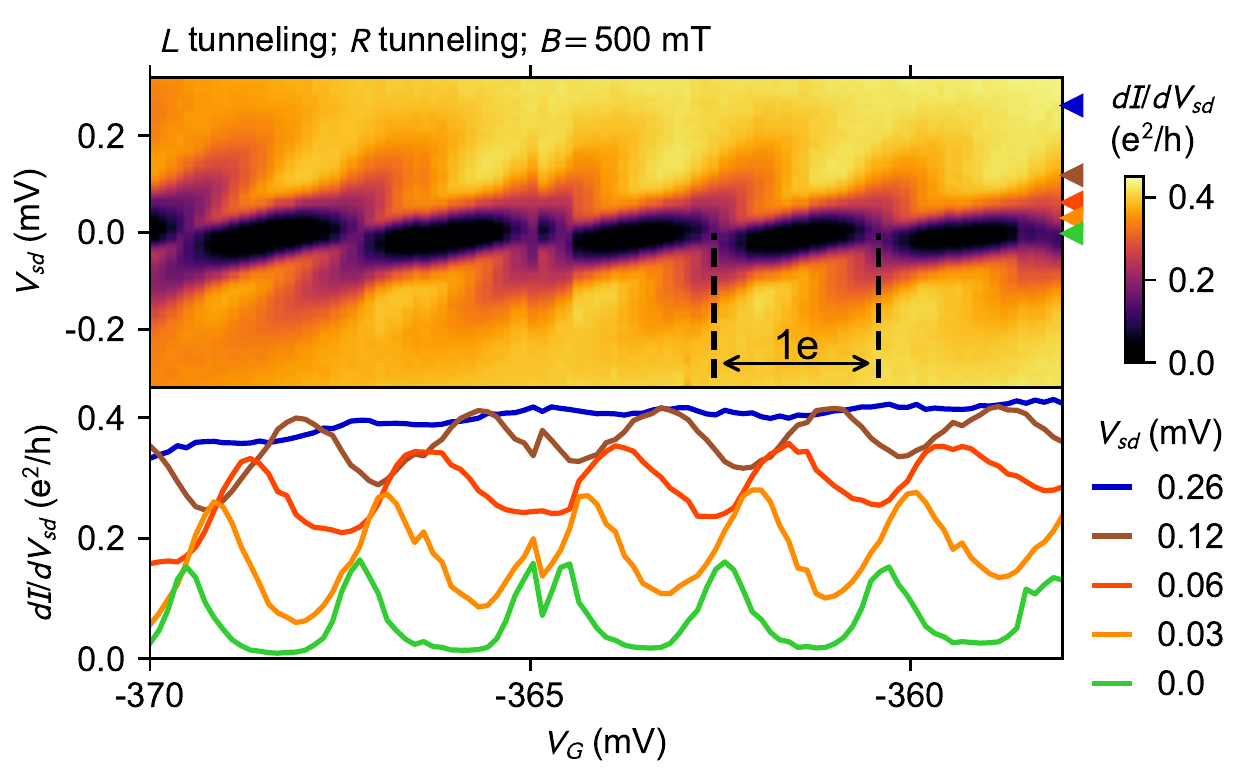}
    \caption{Coulomb diamonds at $B = 500$~mT, in plane of the sample, not parallel to the nanowire obtained by lockin techniques. No conductance suppression occurs at low $V_{sd}$, as exemplified with the cuts presented in the bottom panel. Data for device A.}
    \label{DC_normal}
\end{figure}

Panels (a) and (b) in Fig.~\ref{DC_superconducting} characterize the superconducting gap of the 1$e$ region and the filter, respectively, by mean of tunneling spectroscopy. The tunneling spectroscopy on the left end of the main island reveals a soft gap. In contrast, the right end of the island shows a hard gap ($\Delta_0$ = 308$\pm$2~$\mu$V). We gain further insight into the properties of the 1$e$ region and the filter region in the Coulomb-diamonds measurements presented in Fig.~\ref{DC_superconducting}(c). Besides the 1$e$ periodicity of the diamonds at high DC bias, a zero bias conductance is strongly suppressed. This is shown by the cuts at different bias values in the bottom panel. The zero-bias suppression of conductance at zero magnetic field stands in contrast to the Coulomb diamonds measurement with a magnetic field of 500~mT (Fig.~\ref{DC_normal}) applied in the plane of the sample, but not along the nanowire, to suppress superconductivity in the nanowire Al shell. In this case, the Coulomb peaks have a significant height even at zero DC bias.

From the measurements presented in Figs.~\ref{DC_superconducting},~\ref{DC_normal} we conclude that at zero magnetic field the left side of the device can exchange both single electrons and (Cooper) pairs with the neighboring lead, while the right side can only exchange electron pairs. In particular the suppression of conductance at zero DC bias in Fig.~\ref{DC_superconducting}(c) can be explained as follows. Zero-bias conductance through the right end of the island can only occur if $N$ and $N+2$ occupancy of the island are degenerate and have the lowest energy. However, this never takes place since at the degeneracy point the $N+1$ occupancy has the lowest energy and can be easily reached by exchanging an electron on the left end of the island. This blockade is lifted in magnetic field (Fig.~\ref{DC_normal}), when aluminum becomes normal and both ends of the island can exchange single electrons with the leads.

\begin{figure}[tb]
	\includegraphics[width=0.48\textwidth]{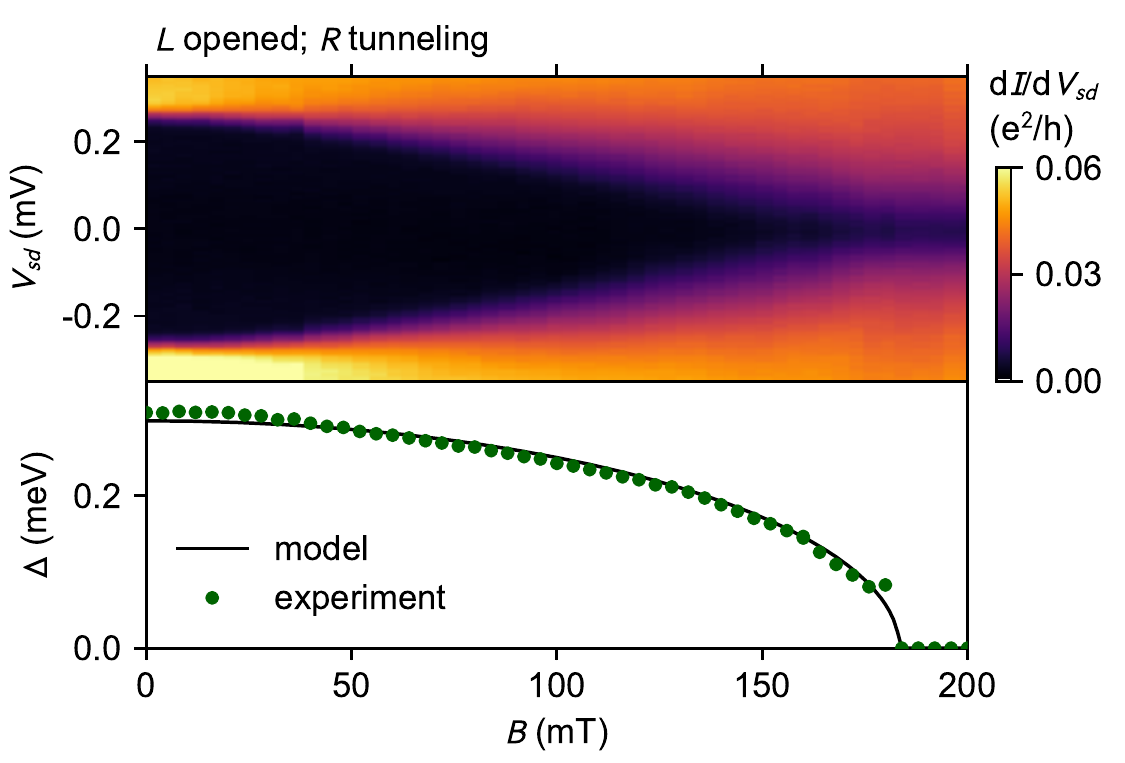}
\caption{Tunneling spectroscopy of the superconducting gap at the filtered end of the main island as a function of external magnetic field $B$ applied approximately 20$^\circ$ away from the wire direction. Bottom panel shows the extracted superconducting gap $\Delta$. The fit of Eq.~\eqref{eq_gap} to the data, presented with a solid line, was used in the further analysis of poisoning rates in the external magnetic field. Data for device A.}
    \label{gap_closing}
\end{figure}

To complete the low frequency characterization of the device, we study the superconducting gap closing on the right side of the device as a function of the magnetic field $B$ (applied in the plane of the device, approximately 20$^\circ$ away from the wire direction); Fig.~\ref{gap_closing}). Panel Fig.~\ref{gap_closing}(b) shows extracted gap size $\Delta$ as a function of $B$. The dependence $\Delta(B)$ is well described by the formula\cite{douglass1961}
\begin{equation}
    \Delta (B) = \Delta_0 \sqrt{1-(B/B_c)^2},
    \label{eq_gap}
\end{equation}
where $\Delta_0 = 308 \pm 2$~$\mu$eV is the fitted superconducting gap at zero field and $B_c=183\pm1$~mT is the fitted critical field (see Appendix \ref{Gap extraction} for more details of the fitting procedure).
These measurement will serve as a reference for study of main island poisoning at finite magnetic field in Section~\ref{efficiency}.

\section{RF charge stability diagram of the island}
\label{RF_charge_diagram}

Next, voltages of the barrier gates are adjusted to study the efficiency of the quasiparticle filter. The voltage of the left barrier gate is set below $V_L = -1$~V to prevent any tunneling into the island (see appendix \ref{Barriers_appendix}) directly into the 1$e$ region (through the barrier \emph{L} in Fig.~\ref{device}(b)). For comparison, the conductance saturates at maximum value for $V_L = 200$~mV and becomes unmeasurable at $V_L = -300$~mV. Meanwhile, the right barrier gate, neighboring the filter, is operated in the tunneling regime, in range $V_R=-80$ to $-250$~mV. The filter gate voltage $V_F$ is set to $-4$~V to create a hard gap in the filter region\cite{vaitiekenas2018}.

\begin{figure}[tb]
	\centering
	\includegraphics[width=0.48\textwidth]{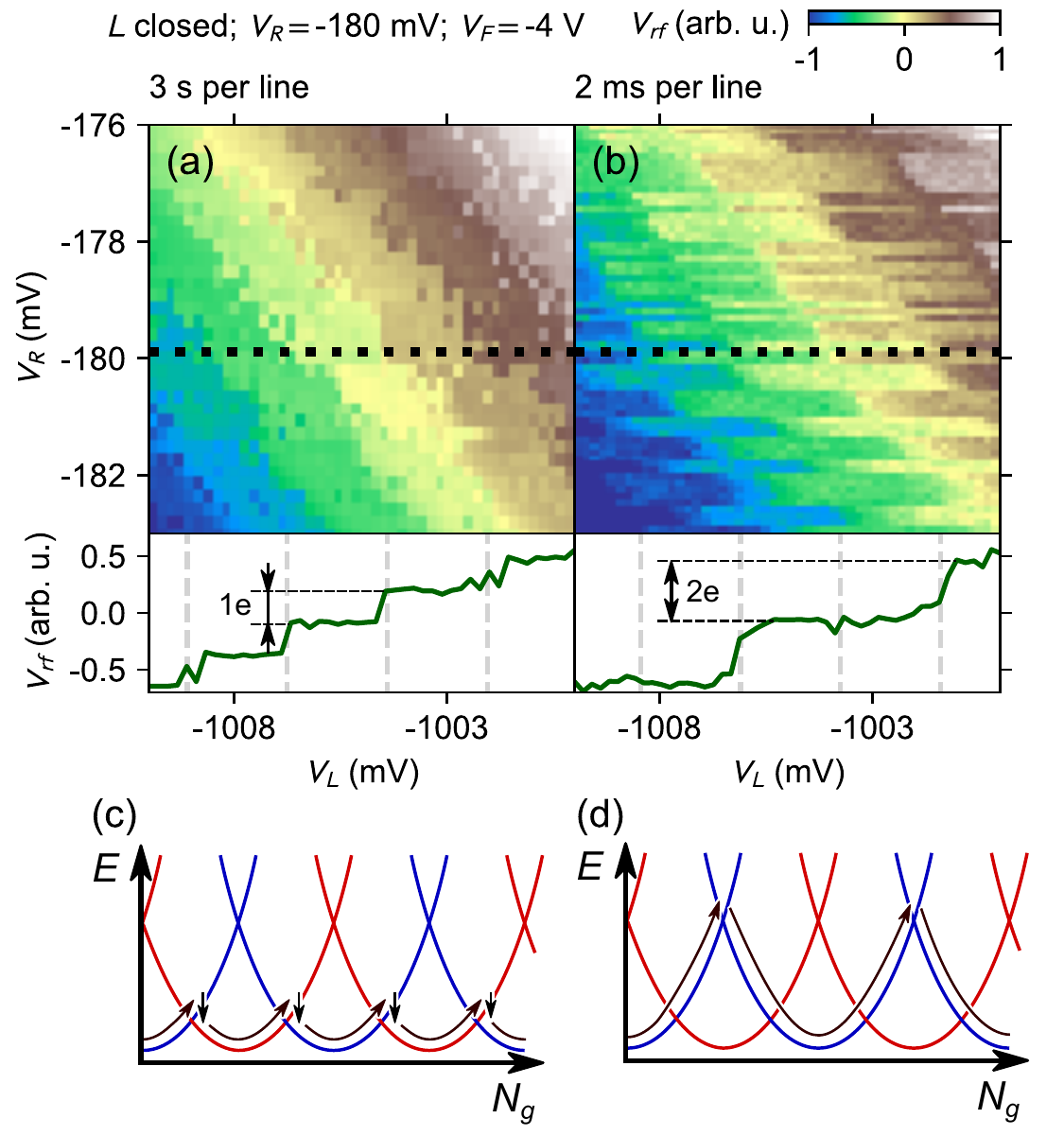}
    \caption{Charge diagrams of the main island with the left barrier fully pinched-off at $V_L \approx -1$~V and right in the tunneling regime at $V_R \approx -180$~mV. Filter gate voltage is set to $V_F=-4$~V. Demodulated rf charge sensing signal was measured with (a) a digital multimeter and averaging time of 20~ms; (b) a waveform digitizer synchronized to 1.903 kHz sawtooth waveform applied to $V_L$, with 4 averages per line. Bottom panels of (a,b) present cuts through the data to illustrate the different periodicity of the charge-sensing staircase. (c,d) Schematic explanation of different observed periodicity of the staircase. The curved arrow follows the energy of the lowest charge state and the downward arrow in (c) represents a poisoning event. In (a,c) the poisoning time is faster than the acquisition rate allowing for the relaxation to the charge state with lowest energy, independently of its parity. In (b,d) the sweeping rate is faster than the poisoning rate an thereby the parity of the island is fixed within each horizontal cut. Data for device B.}
    \label{charge_diagrams}
\end{figure}

Figures~\ref{charge_diagrams}(a,b) present two charge diagrams of the main device as a function of two barrier gate voltages, swept in the same range but with different acquisition rates. The range within which the barrier gate voltages are changed does not significantly affect tunneling rates between the island or any of the leads. Fig.~\ref{charge_diagrams}(a) presents the rf charge-sensing signal measured with a digital multimeter and an integration time of 20 ms. The panel below the charge diagram shows a cut through the data, revealing a characteristic staircase shape\cite{likharev1999,duncan1999}. Each step indicates a charge degeneracy point at which a single electron is added or removed from the island, while plateaus indicate that the charge is fixed.

Figure~\ref{charge_diagrams}(b) shows the same meaurement taken at a faster rate. We apply a 1.903 kHz sawtooth waveform to $V_L$ and measure the charge sensing signal using synchronized waveform digitizer~\cite{stehlik2015}. Each row in the data is an average over 4 periods of the sawtooth. In this way, every row is acquired within 2~ms and averaging time per point is 22~$\mu$s. An example of a single row (cut through the data) is shown in the bottom panel. With the increased acquisition rate, we observe irregular switching between two staircase shapes offset by half a period relative to each other and with twice the period of the staircase in Fig.~\ref{charge_diagrams}(a) (as illustrated with dashed vertical lines).

The contrast between these two datasets makes evident that the poisoning occurs at the timescale between the acquisition time of the data in panels (a) and (b), i.e., tens of milliseconds (corresponding to switching rate of hundreds of $s^{-1}$). Meanwhile the doubled periodicity of the staircase in Fig.~\ref{charge_diagrams}(b) establishes that the tunneling rate of the electron pairs remains much larger than the sawtooth frequency of 1.903 kHz. We dedicate the remainder of the paper to quantifying 2$e$-switching and 1$e$-switching (poisoning) rates as a function of various control knobs. Large tunability of these rates and orders-of-magnitude difference between them are the key findings of this work.

\section{Quantifying poisoning rate}
\label{method}

\begin{figure}[tb]
	\includegraphics[width=0.48\textwidth]{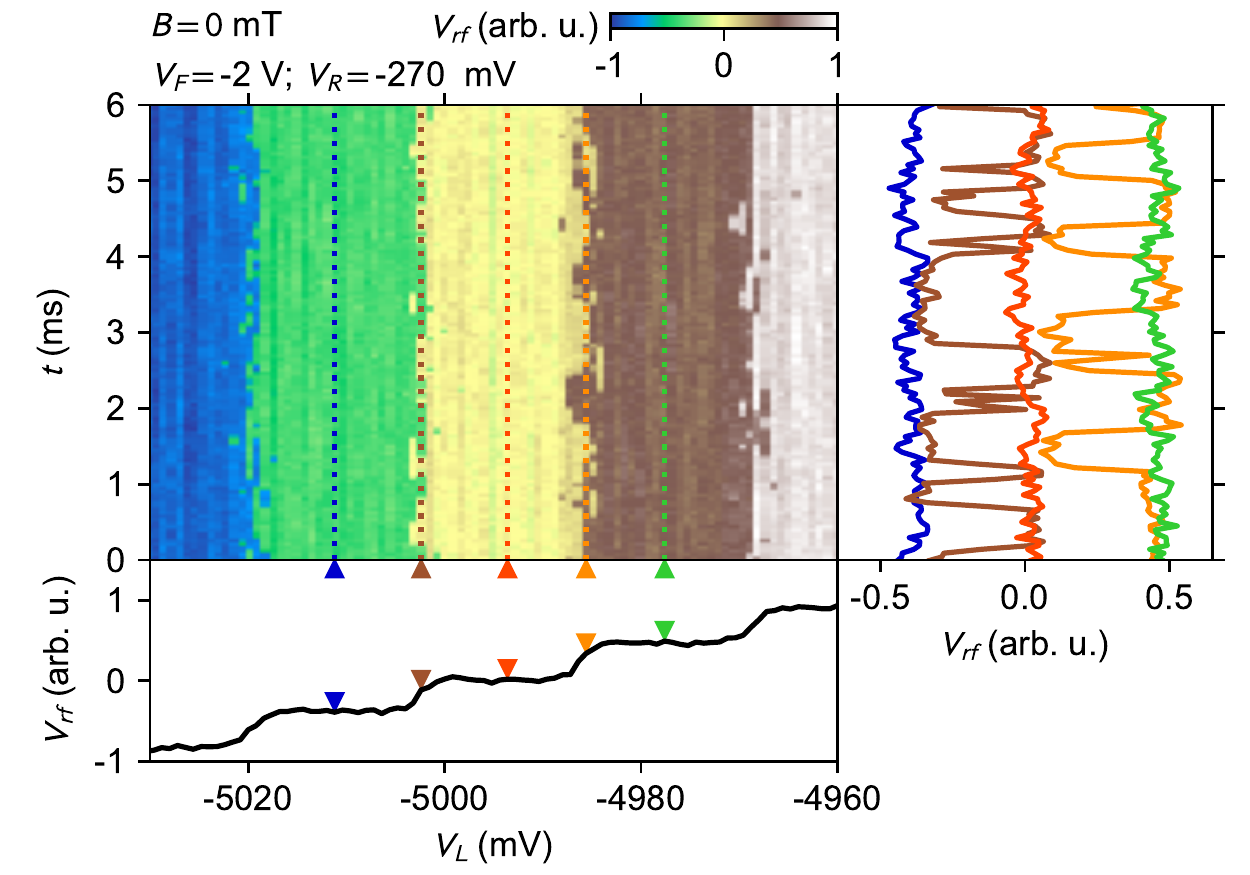}
    \caption{Example of time traces used in the analysis of the switching rate. The two-dimensional color map shows the time vs gate charge signal corrected from its slope (caused by the slope of the sensor's Coulomb peak) to reveal individual charge steps. The bottom panel shows the gate dependence of time-averaged rf charge sensing signal. The right panel shows selected time traces extracted from the main two-dimensional color plot, for values of $V_L$ indicated with dotted lines and triangular markers. Data for device B.}
    \label{switching}
\end{figure}

To quantify the poisoning rate and 2$e$-switching rate the voltage on the filter gate $V_F$ and right barrier gate $V_R$ was fixed while keeping the left barrier gate $V_L$ at a large negative voltage, ranging between $-1$ and $-5$~V. Next, $V_L$ was varied within a small range over which the occupancy of the main quantum dot changed by only a few electrons, thus using the barrier gate as a plunger in that instance. For each value, we acquired a time trace of the signal from the rf charge sensor\cite{HQN2018} as illustrated in Fig.~\ref{switching}, which shows 6-ms long time traces for $V_F=-2$~V and $V_R=-270$~mV. For most values of $V_F$ the occupancy of the dot was stable, with the exception of the vicinity of charge transitions, which showed 1$e$ switching. Time traces obtained at the charge transition and in the stable region are presented in the right panel of Fig.~\ref{switching}. The corresponding values of $V_L$ are indicated with triangular markers in the bottom panel. 

\begin{figure}[tb]
	\includegraphics[width=0.48\textwidth]{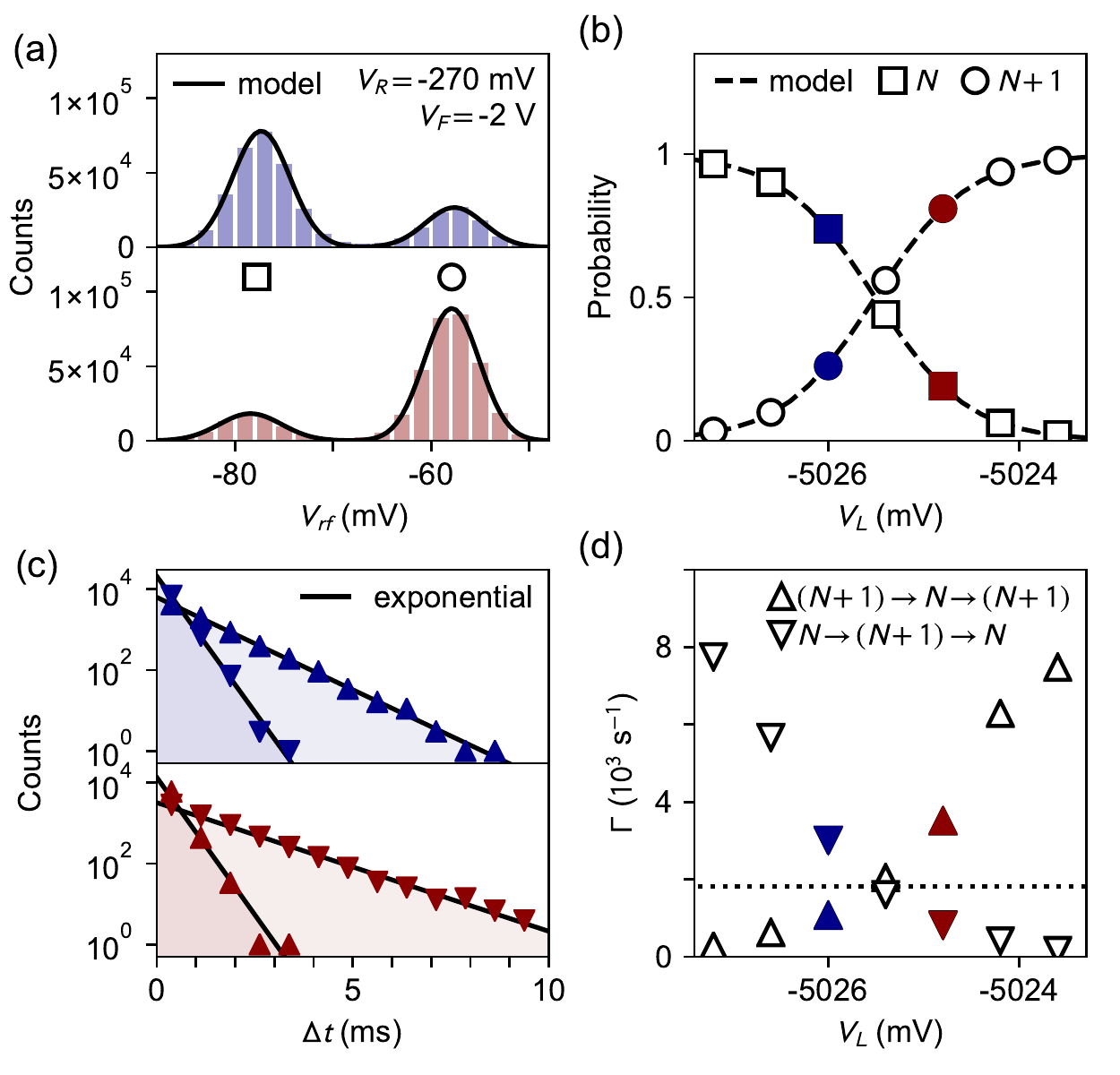}
    \caption{(a) Histograms of rf charge sensing signal for $V_R=-270$~mV, $V_F=-2$~V and $B=0$~mT at two different values of $V_L$ (in blue, before the charge transition and in red, after the charge transition). (b) Probability of finding an island in certain charge state in the vicinity of the charge transition. Solid and dashed lines are fit of the Eq.~\eqref{charge_transition_shape} to the data. (c) Histograms of the waiting times $\Delta t$ between subsequent poisoning events. Triangles pointing up/down indicate electrons tunneling into/out of the island. Solid lines are exponential fits to the data. (d) Switching rates in the vicinity of the charge transition extracted from the exponential fits as in (c). Dotted line indicates $\Gamma=\Gamma_{\rm in}(V_L)=\Gamma_{\rm out}(V_L)$. Colored markers in (b,d) indicate $V_L$ corresponding to data sets in (a,c). Data for device B.}
    \label{analysis}
\end{figure}

We focus on voltages $V_L$ in the vicinity of the charge transition and analyze the corresponding time traces in two ways. First, for each voltage $V_F$ we generate a histogram of the rf charge sensor measurements. The resulting distribution is typically bimodal, as shown in Fig.~\ref{analysis}(a), and well fit by a double gaussian. The two modes correspond to two occupancies of the main island differing by one electron. As demonstrated in Fig.~\ref{analysis}(b) the counts shift between the two modes across a charge transition. We model the probability of detecting charge states $N$ and $N+1$ by a thermal occupation of the island 
\begin{align}
	P_{N} &= \frac{e^{-(N - N_g)^2 E_c/k_B T}}{e^{-(N - N_g)^2 E_c/k_B T} + e^{-(N+1 - N_g)^2 E_c/k_B T}}, \nonumber \\
    P_{N+1} &= 1-P_N,
    \label{charge_transition_shape}
\end{align}
where $N_g = \alpha(V_L-V_L^0)$ is a gate-induced charge, with lever arm $\alpha$ and $V_L^0$ voltage offset; $E_c$ is the island charging energy, $k_B$ is the Boltzmann constant and $T$ is the electron temperature. The two parameters fit ($V_L^0 = 5025.509\pm 0.005$~mV and $\alpha E_c/k_B T = 0.486 \pm 0.005$ mV$^{-1}$) results in a very good agreement with the data.

We now study the characteristic poisoning time. Fig.~\ref{analysis}(c) presents the histogram of time differences between consecutive switches. In both cases the distribution of time between switching events is exponential. In this way, statistics of switching for a specific $V_L$ can be fully characterized by a rate of the electron tunneling in ($\Gamma_{\rm in}$) and out ($\Gamma_{\rm out}$).
Figure~\ref{analysis}(d) presents the change of the tunneling rates across the charge transition. To analyze the filter efficiency we need to exclude poisoning protection by charging energy\cite{HQN2018}. As a poisoning measure we choose a switching rate at gate voltage $V_L$ for which $\Gamma_{\rm in} = \Gamma_{\rm out} \equiv \Gamma$, indicated by a dotted line in Fig.~\ref{analysis}(d). The analysis described in this section underlies each of the $\Gamma$ measurements in the following discussion.

\section{Filter efficiency}
\label{efficiency}

\begin{figure}[tb]
	\includegraphics[width=0.48\textwidth]{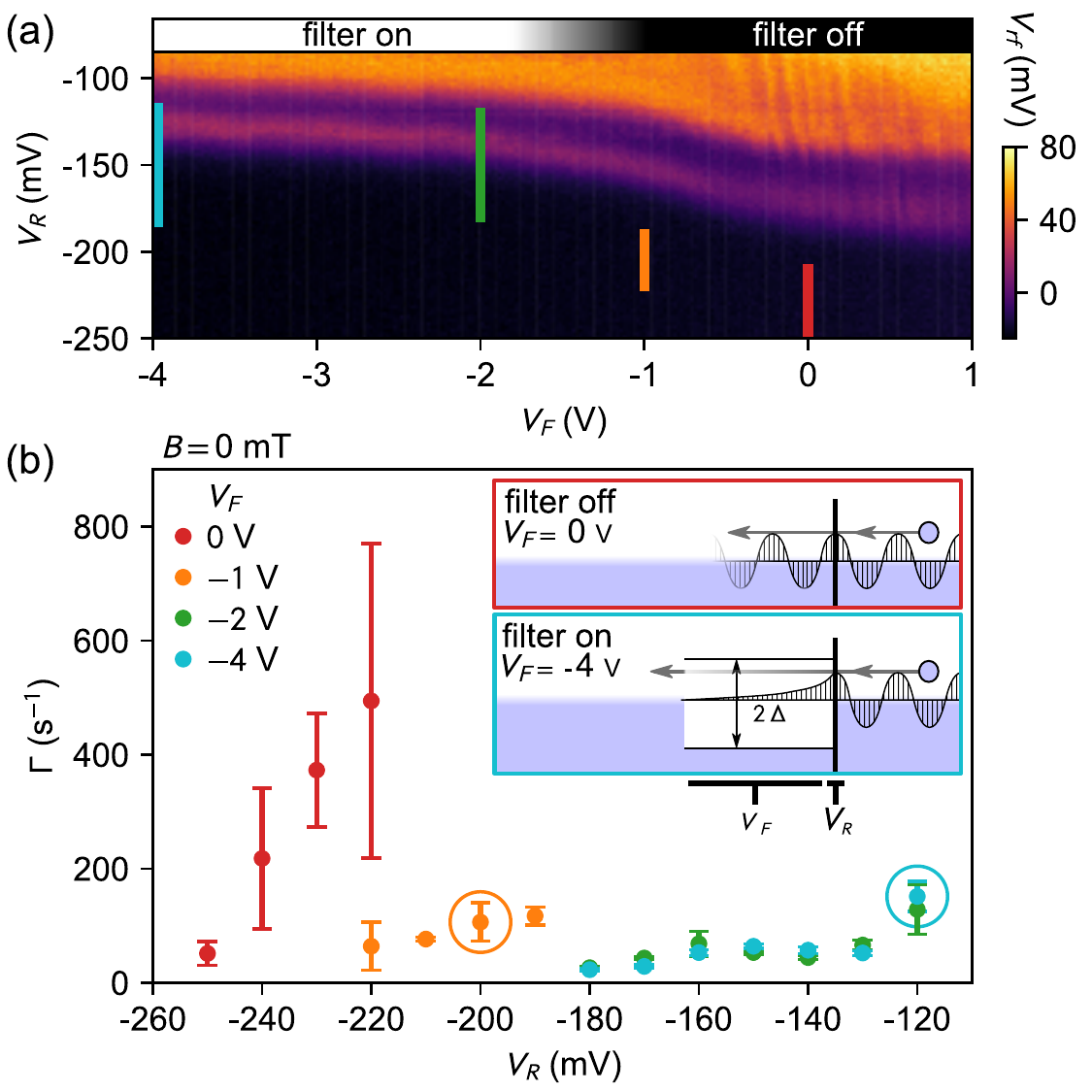}
    \caption{(a) Differential conductance through the right barrier of the main island, with the left barrier fully opened, as a function of $V_R$ and $V_F$ at $V_{sd}=0$, measured by means of rf reflectometry. (b) Dependence of the poisoning rate on $V_R$ for various filter gate voltages $V_F$ and for fully closed $V_L$. The corresponding regions in gate voltage space are indicated by vertical lines in (a). The inset shows the physical mechanism of quasiparticle filtering. When the filter is off (red-framed inset) an incident electron from the right lead needs only to tunnel through the barrier to poison the island. When the filter is on (blue-framed inset) an electron needs to tunnel through both: barrier and 1~$\mu$m-long superconducting region with gap $\Delta$. The raw data behind the points indicated with circles of respective color was analyzed also with an alternative method, as described in Sec.~\ref{fluctuations}. Data for device B.}
    \label{pincher-plunger}
\end{figure}

Having established a quantitative measure of the poisoning rate, we study its dependence on the filter gate $V_F$ and the right barrier gate voltage $V_R$. One should note that the voltages indicated for the different gates at a given conductance value can vary between measurements due to hysteresis, as explained more thoroughly in figure \ref{hysteresis} in the Appendix. We correlate poisoning rates with conventional zero-bias DC measurements of NIS junction at the position of right barrier gate. Figure ~\ref{pincher-plunger}(a) presents the conductance through the NIS junction as a function of $V_F$ and $V_R$, measured by rf reflectometry. For most negative $V_F$, between $-4$ to $-1.5$~V, there is a weak dependence on $V_F$, aside from a weak cross-talk effect resulting in a small negative slope. For $V_F$ between $-1.5$ and 0~V, we observe a smooth step, lowering the voltage for a complete pinch-off using $V_R$ by approximately 50~mV. For $V_F$ above 0~V, the right barrier voltage must be significantly more negative to fully block conductance through the NIS junction.
The nearly vertical features seen at more positive $V_F \approx 0$ can be attributed to an unintentional quantum dot in the barrier\cite{deng2016,lai2019}.

Our interpretation of the data between $V_F = -4$ and $-1$~V is that the semiconducting part of the wire below the filter gate is maximally depleted, and strongly coupled to the superconductor\cite{chang2015,vaitiekenas2018,mikkelsen2018,antipov2018,winkler2018}. For more positive $V_F$, carriers are accumulated in the semiconductor.

We find that quasiparticle tunneling rates agree with this interpretation. Fig.~\ref{pincher-plunger}(b) presents the tunneling rates $\Gamma$ obtained at the charge degeneracy points for several values of $V_F$ and $V_R$. Colored lines in Figure~\ref{pincher-plunger}(a) indicate corresponding positions in the color map. For all values of $V_F$ tunneling rates decrease with decreasing voltage $V_R$. This dependence can be easily understood considering that the barrier gate voltage $V_R$ changes the transparency of the NIS junction and suppresses quasiparticle poisoning. However, this suppression also reduces the probability of Andreev reflections and does not result in quasiparticle-filtering properties. On the other hand, decreasing the filter gate voltage $V_F$ disproportionately affects 1$e$ tunneling rate. As demonstrated in Fig.~\ref{pincher-plunger}(b), the single electron tunneling rate for $V_F =-2$~V is below 200 s$^{-1}$ in the entire studied range of $V_R$, between $-$180 and $-$120~mV (for more positive $V_R$ the charge sensing signal disappears due to strong 2$e$ quantum fluctuations, see Sec. \ref{fluctuations}). To achieve the same tunneling rate with $V_F = 0$, $V_R$ has to be decreased to $-250$~mV. As illustrated with colored lines in Fig.~\ref{pincher-plunger}(a), this is a disproportional change relative to the 50~mV shift of the pinch-off curve between $V_F=-2$ and 0. At the same time, a decrease of $V_F$ from $-2$ to $-4$~V has virtually no impact on the poisoning rates, consistent with reaching a maximum in the proximity coupling of semiconducting states \cite{mikkelsen2018,antipov2018,winkler2018}.

Our interpretation of these observations is illustrated in the insets of Fig.~\ref{pincher-plunger}(b). When the voltage {$V_F  < -2$~V}, the semiconductor of the gated nanowire segment is depleted sufficiently to have a hard gap and in order to poison the island quasiparticles must tunnel through a 1~$\mu$m long region with gap $\Delta$ (blue-framed inset). On the other hand, when $V_F\geq$ 0 the semiconductor is not depleted and quasiparticles only have to tunnel through the barrier junction (red-framed inset). 

\begin{figure}[tb]
	\includegraphics[width=0.48\textwidth]{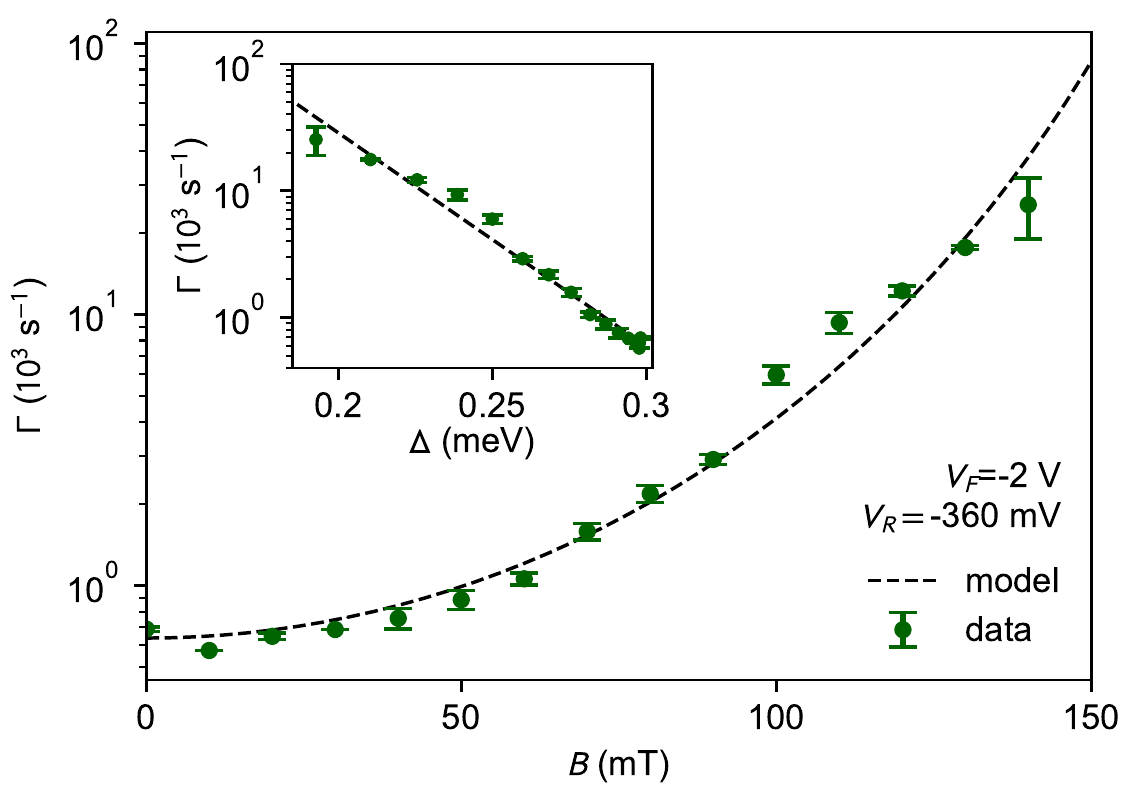}
    \caption{Field dependence of the poisoning rate for $V_F=-2$~V, $V_R=-360$~mV. The dahsed line is a model fit to the data using the field dependence of the superconducting gap $\Delta(B)\propto 1/\xi(B)$ fitted to the data in Fig.~\ref{gap_closing}(a). For description of the model see the main text. The inset shows the poisoning rates as a function of superconducting gap size $\Delta(B)$. Data for device A.}
     \label{B_sweep}
\end{figure}

Next we study whether applying an external magnetic field influences the poisoning rate. Figure~\ref{B_sweep} shows the magnetic field dependence of the poisoning rate for $V_F=-2$~V and $V_R=-360$~mV in device A (different from the one discussed in Fig. \ref{pincher-plunger}). The poisoning rate increases as the magnetic field that suppresses the gap of the filter.

We model poisoning processes as limited by single-particle tunneling through the filter, 
\begin{equation}
	\Gamma = Ae^{-L/\xi(B)},
\end{equation}
where $A$ is the tunneling rate excluding the superconducting filter and $L=1$~$\mu$m is the length of the quasiparticle filter. The magnetic field dependence of the coherence length $\xi(B)$ can be modeled within BCS theory as $\xi(B)=\hbar v_F/\pi\Delta(B)$ where $v_F$ is the Fermi velocity and $\Delta(B)$ is obtained from the DC measurement in Fig.~\ref{gap_closing}(a). Specifically we use the relation $L/\xi(B)=\Delta(B)/b$, where $b$ is a fit parameter (with $b=v_F/\pi L$ for clean case in BCS theory). Using this model and data from Fig.~\ref{gap_closing}(b) we fit the field dependence of $\Gamma$ in Fig. \ref{B_sweep}. We find that $A=66\pm28 \times 10^6$~s$^{-1}$ and $b=26\pm1$~$\mu$eV yields a good matching with the data for poisoning rates ranging two orders of magnitude. The value of $b$ is significantly larger than $k_B T \approx 4$~$\mu$eV, which suggests that the field-dependence of the tunneling rate cannot be explained by thermally activated quasiparticles crossing the filter but rather by quantum tunneling. The corresponding coherence length is $\xi(0) = 90 \pm 10$~nm, taking into account a 50~nm uncertainty in the length of the filter (accounting for the size of the barrier gates). This value of $\xi$ at zero magnetic field is one order of magnitude smaller that that of bulk Al (1600~nm) and a factor of 2-3 smaller than the inferred coherence length in topological regime in Ref. \cite{albrecht2016} and Ref.{albrecht2017} (260~nm), and Ref. \onlinecite{vaitiekenas2018_2} (180~nm).

\section{Quantum charge fluctuations in presence of quasiparticle filtering}
\label{fluctuations}

In this section, we demonstrate that the quasiparticle filter allows maintaining a low poisoning rate while simultaneously having strong coupling to the lead, as indicated by transition broadening due to 2$e$ quantum charge fluctuations.
For that purpose, we reanalyze two of the data sets, adjusted to have comparable poisoning rates but corresponding to different settings of the barrier gate voltage $V_R$. To achieve comparable poisoning rates, the data set in the regime of active filter (filter on) is operated with the right cutter in the open regime while for the other data set, in the filter-off regime, the right cutter is almost closed. The two parameter regimes are indicated by the encircled data points in Fig.~\ref{pincher-plunger}.

\begin{figure}[tb]
	\includegraphics[width=0.48\textwidth]{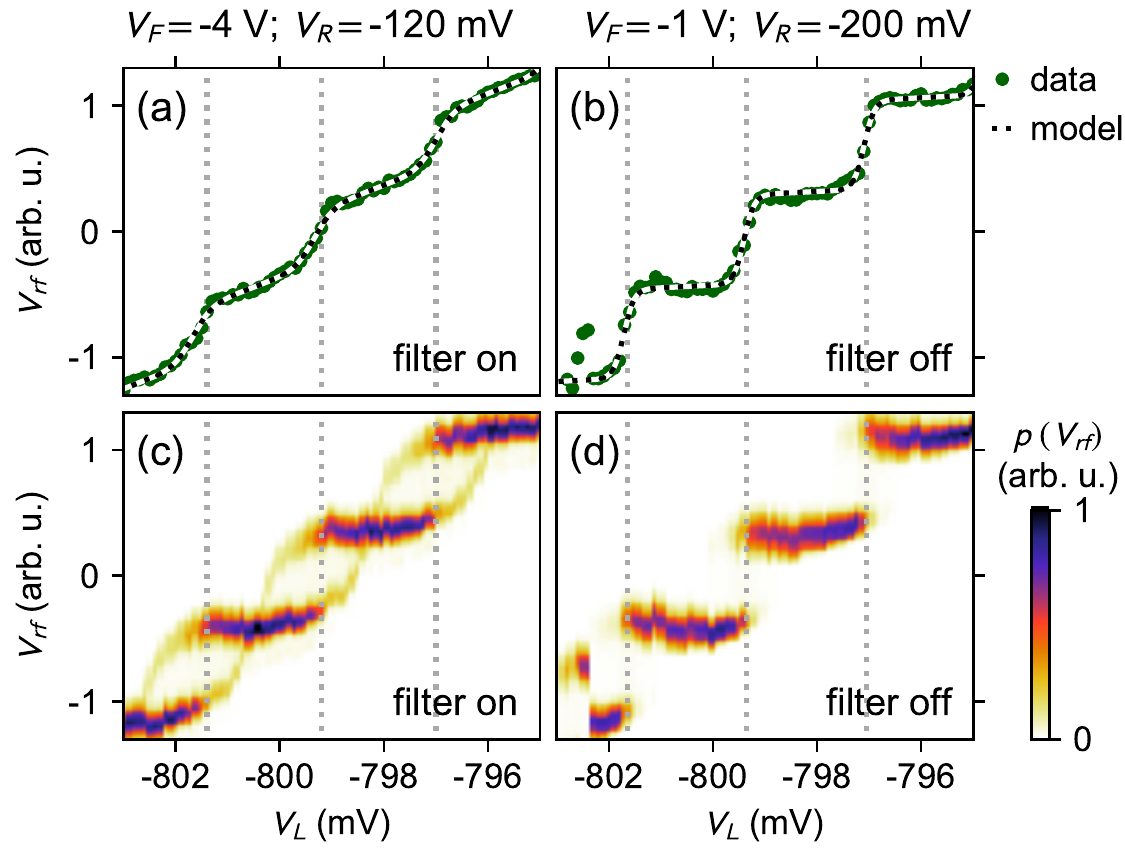}
    \caption{Charge steps measured with quasiparticle filter on (a,c) and off (b,d). Panels (a,b) show the averaged time traces, as in Fig.~\ref{switching}, fitted by formula in Eq.~\eqref{q-thermal}. Panels (c,d) show histograms of charge sensing signal, as in Fig.~\ref{analysis}(a), as a function of $V_L$. Data for device B.}
    \label{histograms}
\end{figure}

Time-averaged charge sensing signal for these two data sets is presented in Fig.~\ref{histograms}(a,b) as a function of $V_L$ with green points. In both cases, we observe a characteristic staircase shape. However, the steps are much more pronounced when the filter is ``off'' relative to the filter ``on'' configuration. In both cases the periodicity corresponds to a 1$e$ charging effects. In the following we test the consistency of the data with an exclusively thermal broadening of the charge transitions. Below we demonstrate that this model holds well in the filter ``off'' case but fails in the filter ``on'' case.

For purely thermal broadening the time averaged charge is given by the thermodynamic expectation value,
\begin{equation}
    \langle Q \rangle = e\times\frac{ \sum\limits_{n}
            n \exp \left(-(n-N_g)^2 E_c/k_B T \right)
            }{\sum\limits_{n}
            \exp \left(-(n-N_g)^2 E_c/k_B T \right)}.
    \label{q-thermal}
\end{equation}
Taking the charge sensor sensitivity and direct crosstalk between gate $V_L$ and the sensor into account, a fit of Eq.~\eqref{q-thermal} to the data yields $E_c / k_B T = 6.3\pm1.0$ and $11.7\pm1.0$ for filter ``on'' and ``off'' case, respectively. The result is plotted in Fig.~\ref{histograms}(a,b) with black dotted lines.

Next, we generate histograms of charge sensor measurements separately for each of the time traces taken at different $V_L$, as in Sec.~\ref{method} and Fig.~\ref{analysis}(a). The histograms for various $V_F$ are plotted as a colormap in Fig.~\ref{histograms}(c,d). In the filter off case the distribution is unimodal, except in the vicinity of the charge transition. We attribute this to the stability of a single charge state whenever Coulomb blockade suppresses poisoning. Meanwhile, when the filter is on, the distribution is always bimodal, corresponding to two distinct charge states. From this we conclude that the two lowest charge states must always differ in energy by not much more than $k_B T$, otherwise the probability of finding the system in the excited state would be negligible. This corroborates the factor-of-two difference in $E_c / k_B T$ obtained from the fit in Figs.~\ref{histograms}(a,b).

More intriguingly, the mode corresponding to the excited state never disappears. Instead, the mode corresponding to a charge state $N$ \emph{continuously} shifts to a position corresponding to a charge $N\pm2$. We consider two alternative explanations of these superimposed and shifted 2$e$ steps. First, it is possible that the 2$e$ tunneling rate exceeds significantly the temporal resolution of the charge sensor while 1$e$ tunneling rate is within our temporal resolution. In such case, we detect the time-averaged charge with fixed parity. Alternatively, it is possible that we observe quantum charge fluctuations, that cause the charge of the island to be non-integer\cite{feigelman2002,lutchyn2016,jezouin2016}.

\begin{figure}[tb]
	\includegraphics[width=0.48\textwidth]{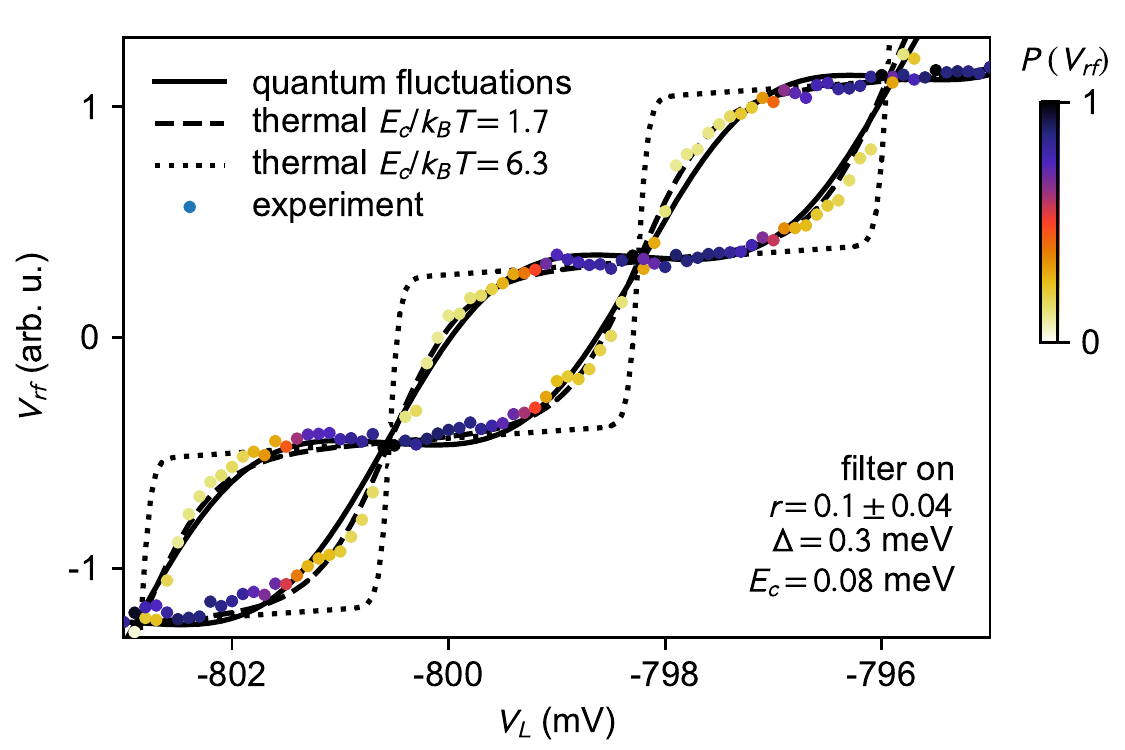}
    \caption{Extracted positions and amplitudes of peaks of histograms presented in Fig.~\ref{histograms}(c). The solid line is the model of quantum charge fluctuations  fitted to the data. Superconducting gap $\Delta=0.3$~meV and charging energy $E_c=0.08$~meV used in the model were measured independently. The simultaneous fit of the two presented curves yields a reflection coefficient of $r = 0.10 \pm 0.04$. Dashed and dotted lines present predictions of the thermal model (see text) for $E_c/k_B T=$1.7 and 6.3, respectively. Data for device B.}
    \label{hybridzation}
\end{figure}

To distinguish between those possibilities we process the data in Fig.~\ref{histograms}(c) by fitting double gaussian distributions to each of the histograms to obtain charge sensing signal corresponding to both detected charge states as well as probability of detecting each of them. The result of this procedure is plotted in Fig.~\ref{hybridzation} and tested against the two hypothesis.

To test the hypothesis of thermal broadening, we fit Eq.~\eqref{q-thermal} using only $n$-even or only $n$-odd, simultaneously to the positions of the two modes. Leaving $E_c/k_B T$ free allows for a good fit to the data, presented with the dashed line in Fig.~\ref{hybridzation}. However, the obtained value of $E_c/k_B T = 1.7 \pm 0.2$ is a factor of 4 smaller than the value $E_c/k_B T = 6.3 \pm 0.7$ obtained earlier \emph{for the same data set} from the fit in Fig.~\ref{histograms}(a). Alternatively, a fixed value of $E_c/k_B T = 6.3$ results in a poor agreement with the data, presented with the dotted line in Fig.~\ref{hybridzation}. We conclude that the observed behavior can not be explained in terms of thermal broadening.

For a description of quantum charge fluctuations we employ the model developed in Ref.~\onlinecite{lutchyn2016} to quantify the coupling strength between the island and the normal lead. The main island of our device is treated as a superconductor coupled to a normal lead. The coupling strength is then quantified in terms of the normal reflection coefficient $r$ at the interface. Assuming that the coupling is to a single-channel, for $r \ll 1$ the ground state energy of the island is then given (to the second order in $r$) by:
\begin{equation}
    E_{GS} = -\sqrt{E_c \Delta r} \cos(\pi N_g) - E_c r^2 \ln \left(\frac{E_c}{\Delta} \right) \cos^2 \left(\pi N_g \right)
\end{equation}
where $\Delta_0$ is the superconducting gap of the filter at $B=0$ and $E_c=0.08$~meV is the bare charging energy of the island (not taking into account the renormalization due to quantum fluctuations). $E_c$ is extracted from several Coulomb diamonds measurements in both normal and superconducting state of the Al shell. This model predicts\cite{lutchyn2016} the mean charge of the island to be
\begin{equation}
    Q(N_g) = e N_g - \frac{e}{2 E_c} \frac{\partial E_{GS}}{\partial N_g}.
\end{equation}

As previously, we supplement this formula with the charge sensor sensitivity and direct crosstalk between gate $V_L$ and the sensor. Next, we perform simultaneous fits of $Q(N_g)$ and $Q(N_g+1)$ to the extracted positions of the two modes. The result is plotted in Fig.~\ref{hybridzation} with a solid line. The fit yields a normal reflection coefficient of $r=0.1 \pm 0.04$, corresponding to dimensionless conductance $g = 1- r^2 = 0.99 \pm 0.01$. This value for the conductance indicates that in the filter ``on'' case, the coupling to the lead is very strong while still keeping very low poisoning rates.

We note that the close to perfect transmission obtained from the single-channel fit indicates that while quantum charge fluctuations are strong the single-channel model likely does apply in our experiment. Instead, the island is rather to be coupled to multiple channels in the normal leads. To the best of our knowledge there is no analytical formula for the expected scenario that includes more than the first order approximation in $1-g$. However, we do not expect the charge steps to take significantly different shapes from the one assumed above.

\section{Conclusions}
\label{conclusions}

To summarize, we studied the poisoning rate of gapless states in a nanowire island that are separated from a quasiparticle reservoir (normal lead) by a hard-gap superconducting segment of the island (filter). By means of radio-frequency charge sensing, we directly probe the poisoning events.

Poisoning rates that are consistent with tunneling through a 1~$\mu$m long (length of our quasiparticle filter) barrier of the height of the induced superconducting gap $\Delta(B)$. The efficiency of the filter is highly tunable by electrostatic gating of the filter section. The single electron tunneling rates can be made lower than $\Gamma=200$~s$^{-1}$. In the same configuration, we observe strong 2$e$ quantum charge fluctuations between the island and the lead.
Moreover, according to our interpretation, the quasiparticle poisoning rate could be exponentially suppressed by increasing the length of the quasiparticle filter.

These results demonstrate that a quasiparticle filter based on the tunability of the superconducting gap has been realized.

\section*{ACKNOWLEDGEMENTS}

We thank S. Upadhyay for contributions to device fabrication and F. Kuemmeth for contributions to the experimental setup. We also thank R. Lutchyn for fruitful discussions.

Research was supported by Microsoft, the Danish National Research Foundation, and the European Commission.

\appendix

\setcounter{figure}{0}
\setcounter{equation}{0}
\renewcommand{\thefigure}{A\arabic{figure}}  
\renewcommand{\theequation}{A\arabic{equation}}

\section{Magnetic field dependence of the superconducting gap}
\label{Gap extraction}

In order to extract the gap from the raw data we perform a fit of each spectrum. This is performed by convolving the Fermi-Dirac distribution at temperature $T$ with the following formula for the raw gap\cite{dynes1978}
\begin{equation}
    \rho(E) = \Re\left(\frac{E+i\gamma}{\sqrt{(E+i\gamma)^2-\Delta^2}}\right),
\end{equation}
where $\gamma$ is a phenomenological parameter describing the finite in-gap conductance. $E$ is the applied voltage bias and $\Delta$ is the superconducting gap we want to extract.

\begin{figure}[t]
	\includegraphics[width=0.48\textwidth]{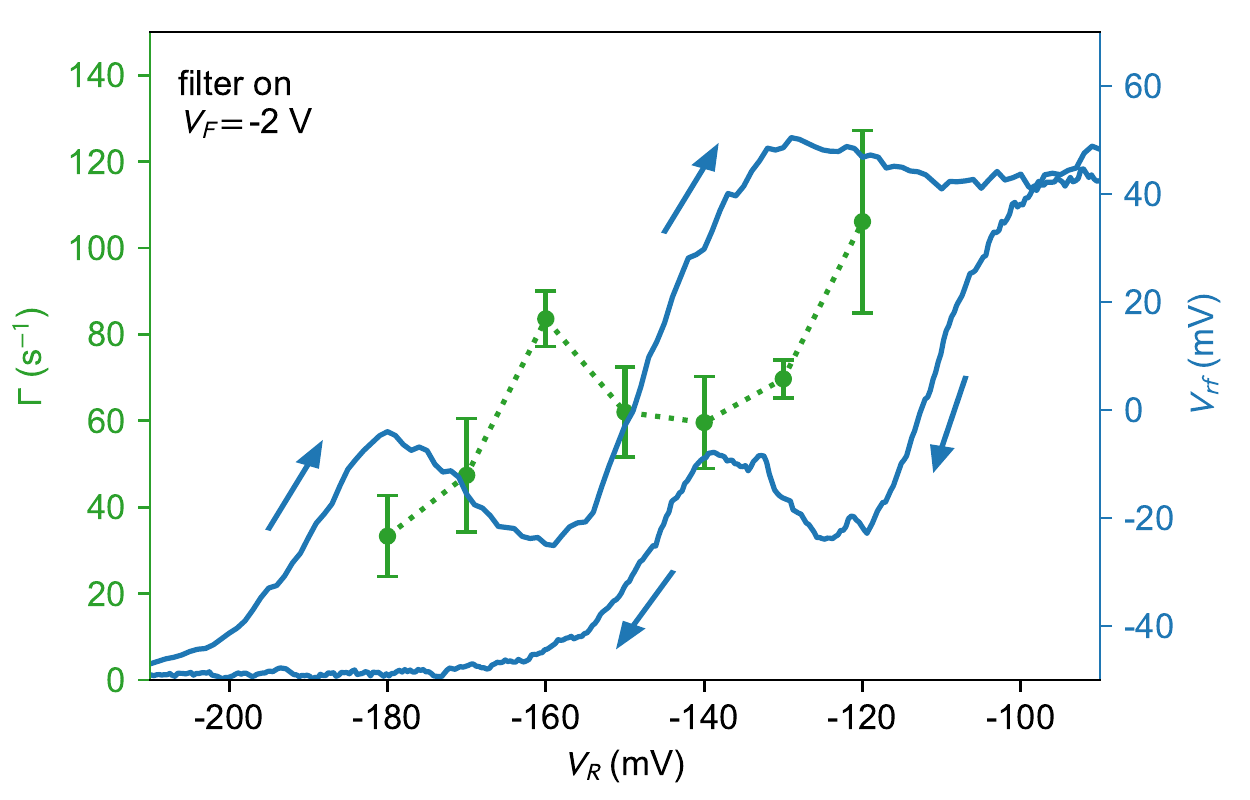}
    \caption{Right barrier characterization for device B. The blue curves show the reflectometry signal in both up and down sweeps of the right barrier. The poisoning rate extracted at $V_F = -2$~V is overlaid as a comparison to the conductance pinch-off.}
    \label{hysteresis}
\end{figure}

In order to perform the fit we fix the temperature $T = 55$~mK and the phenomenological parameter $\gamma = 0.01$~meV as extracted from the zero field data. These values are obtained together with a value of the gap at zero field $\Delta_0 = 0.3$~meV. We then proceed in fitting the gap with these fixed parameters at various values of the magnetic field. The resulting values of $\Delta(B)$ are plotted with dots on Fig. ~\ref{gap_closing}(b). In order to eliminate the fluctuations and the non-monotonicity inherent to that type of procedure we then fit the extracted $\Delta(B)$ dependence to the BCS gap from Eq.~\eqref{eq_gap} leading to a critical field of $183\pm1$~mT.

\setcounter{figure}{0}
\setcounter{equation}{0}
\renewcommand{\thefigure}{B\arabic{figure}}  
\renewcommand{\theequation}{B\arabic{equation}}

\section{Barriers characterization}
\label{Barriers_appendix}

\begin{figure}[tb]
	\includegraphics[width=0.48\textwidth]{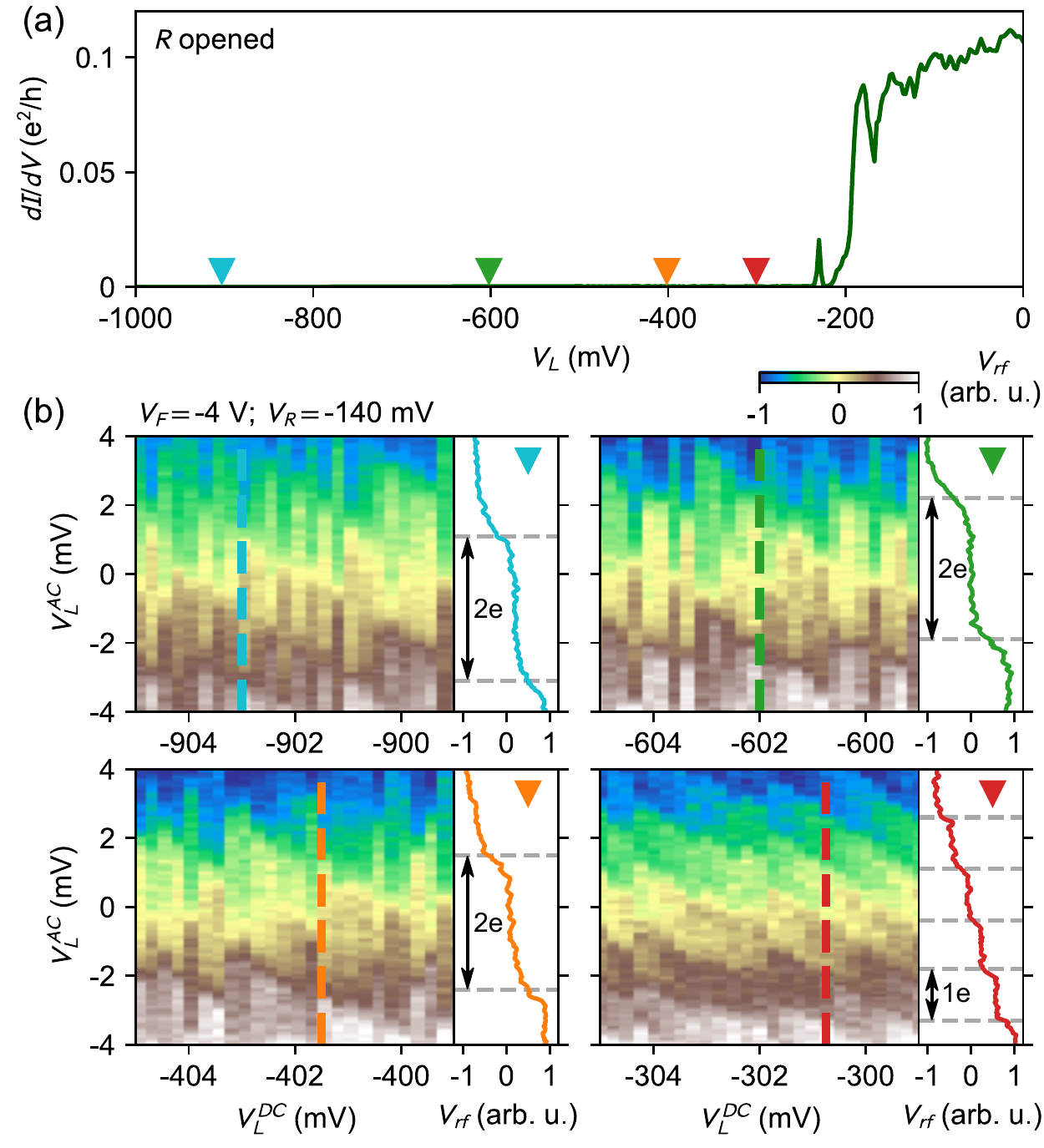}
    \caption{Left barrier characterization for device B. Panel (a) shows the pinch-off curve of the left barrier in the range 0 to $-1$~V obtained by lockin measurement while keeping the right side completely open. Panels in (b) show the charge steps acquired in reflectometry while varying the value of the left barrier and keeping it closed with respect to the pinchoff criterion. The x axis correspond to slow sweeping of the barrier voltage while the y axis corresponds to an AC sawtooth excitation of the same gate. The barrier values at which these data sets were taken are indicated in (a) by colored triangle corresponding to the color of the side panels. These side panels show the typical charge steps taken in each of the configurations on which one can see the evolution from 1$e$ charge steps to 2$e$ charge steps as the barrier is turned more open.}
    \label{left_barrier}
\end{figure}

To supplement our discussion about the relation between the pinch-off of the tunneling barriers and the poisoning rate we present in Fig. \ref{hysteresis} the up and down pinch-off curve from the right barrier. The pinch-off point of the conductance through the right barrier is obtained for a voltage $V_R\simeq -170$~mV when sweeping from positive to negative voltages. When sweeping in the opposite direction the pinch-off can be obtained around a voltage $V_R\simeq -210$~mV. The hysteresis for the pinch-off of the right barrier is thus of approximately $\Delta V_R = 40$~mV. On top of these two curves we superpose the poisoning rate obtained in the same filter on configuration at $V_F = -2$~V. A correlation between the pinch-off curves and the poisoning rate can be observed. However, due to hysteresis of the tunneling barriers (which causes the offset between the poisoning curve and the pinchoff curves) we were not able to simultaneously record the conductance signal and the poisoning rate. This complements our interpretation from the main text by suggesting that the the poisoning rate is directly dependant on the value of the tunneling barrier.

We show in Fig. \ref{left_barrier}(a) the pinch-off curve obtained on the left barrier while having the right barrier completely open ($V_R$ = 1~V). The pinch-off point in obtained around $V_L\simeq -220$~mV. However, despite losing all transport signal in this configuration, we show in Fig. \ref{left_barrier}(b) that the barrier value can still play a role in the charge transfer rate through the left side when the system is placed in a Coulomb blockade configuration. In these figures, the right side is kept at a voltage of $-140$~mV corresponding to a near complete pinch-off as seen from Fig. \ref{hysteresis}.

When the left barrier is at its lowest voltage (blue triangle) the first panel of Fig. \ref{left_barrier}(b) shows that the Coulomb stairs are 2$e$ periodic. This indicates that the poisoned part of the island is not able to exchange single electron on the time scale of one sweep and that only paired electrons can penetrate the island through the right lead. This situation persist until the fourth panel (red) where we recover a situation of 1$e$ periodicity of the charge steps. Such behavior is once again consistent with a poisoned side of the island by means of the NbTiN patch.

\bibliography{biblio.bib}

\end{document}